\documentclass[sigconf]{acmart}
\makeatletter
\def\@ACM@checkaffil{
    \if@ACM@instpresent\else
    \ClassWarningNoLine{\@classname}{No institution present for an affiliation}%
    \fi
    \if@ACM@citypresent\else
    \ClassWarningNoLine{\@classname}{No city present for an affiliation}%
    \fi
    \if@ACM@countrypresent\else
        \ClassWarningNoLine{\@classname}{No country present for an affiliation}%
    \fi
}
\makeatother

\usepackage{adjustbox}
\usepackage{amsmath}
\usepackage[ruled,vlined,linesnumbered]{algorithm2e}
\usepackage{booktabs} 
\usepackage{balance}
\usepackage{chronology}
\usepackage{color}
\usepackage{colortbl}
\usepackage{comment}
\usepackage{etoolbox}
\usepackage{enumitem}
\usepackage{listings}
\usepackage{multirow}
\usepackage{natbib}
\setcitestyle{square, comma, numbers,sort&compress}
\usepackage{pifont}
\usepackage{seqsplit}
\usepackage{subcaption}
\usepackage{tikz}
\usepackage{xspace}
\newcommand{\ie}{i.e.\xspace}
\newcommand{\etal}{\textit{et. al}\xspace}
\newcommand{\eg}{e.g.\xspace}

\newcommand{\todo}[1]{\textcolor{blue}{#1}}
\newcommand{\change}[1]{{#1}}

\newcommand{\apktool}{\textsc{ApkTool}\xspace}

\newcommand{\virustotal}{\textsc{VirusTotal}\xspace}
\newcommand{\drebin}{\textsc{Drebin}\xspace}
\newcommand{\threatbook}{\textsc{ThreatBook}\xspace}
\newcommand{\virusshare}{\textsc{VirusShare}\xspace}
\newcommand{\genome}{\textsc{Genome}\xspace}
\newcommand\xiu[1]{{#1}}

\definecolor{pblue}{rgb}{0.13,0.13,1}
\definecolor{pgreen}{rgb}{0,0.5,0}
\definecolor{pred}{rgb}{0.9,0,0}
\definecolor{pgray}{rgb}{0.46,0.45,0.48}
\definecolor{ppurple}{rgb}{1,0.2,1}
\definecolor{pblack}{rgb}{0,0,0}
\newcounter{finding}
\newenvironment{finding}{
	\refstepcounter{finding}\par\smallskip
	\noindent \textbf{Finding~\thefinding.~}\itshape
}
{
	\par\smallskip \normalfont
}
\AtBeginDocument{%
  }

\setcopyright{acmlicensed}
\copyrightyear{2025}
\acmYear{2025}
\acmDOI{10.1201/9781003121510-2}
\acmConference[Conference acronym 'Internetware2025]{the 16th International Conference on Internetware}{June 20--22,
  2025}{Trondheim, Norway}
\acmISBN{xxx}

\begin{document}

\title{Measuring and Explaining the Effects of Android App Transformations in Online Malware Detection}

\author{Guozhu Meng}
\affiliation{%
  \institution{Institute of Information Engineering, Chinese Academy of Sciences}
  \institution{School of Cyber Security, University of Chinese Academy of Sciences}
}

\author{Zhixiu Guo}
\affiliation{%
  \institution{Institute of Information Engineering, Chinese Academy of Sciences}
  \institution{School of Cyber Security, University of Chinese Academy of Sciences}
}

\author{Xiaodong Zhang}
\affiliation{%
  \institution{Institute of Information Engineering, Chinese Academy of Sciences}
  \institution{School of Cyber Security, University of Chinese Academy of Sciences}
}
\author{Haoyu Wang}
\affiliation{%
  \institution{School of Cyber Science and Engineering,
Huazhong University of Science and Technology}
}
\author{Kai Chen}
\affiliation{%
  \institution{Institute of Information Engineering, Chinese Academy of Sciences}
  \institution{School of Cyber Security, University of Chinese Academy of Sciences}
}

\author{Yang Liu}
\affiliation{%
  \institution{College of Computing and Data Science, Nanyang Technological University, Singapore}
}


\renewcommand{\shortauthors}{Meng et al.}


\begin{abstract}
It is well known that antivirus engines are vulnerable to evasion techniques (\eg, obfuscation) that transform malware into its variants. However, it cannot be necessarily attributed to the effectiveness of these evasions, and the limits of engines may also make this unsatisfactory result. 
In this study, we propose a data-driven approach to measure the effect of app transformations to malware detection, and further explain why the detection result is produced by these engines. 
First, we develop an interaction model for antivirus engines, illustrating how they respond with different detection results in terms of varying inputs. 
Six app transformation techniques are implemented in order to generate a large number of Android apps with traceable changes. 
Then we undertake a one-month tracking of app detection results from multiple antivirus engines, through which we obtain over 971K detection reports from \virustotal for 179K apps in total. 
Last, we conduct a comprehensive analysis of antivirus engines based on these reports from the perspectives of signature-based, static analysis-based, and dynamic analysis-based detection techniques. 
The results, together with 7 highlighted findings, identify a number of sealed working mechanisms occurring inside antivirus engines and what are the indicators of compromise in apps during malware detection. 
\end{abstract}



\maketitle

\section{Introduction}\label{sec:intro}

Commercial antivirus engines (referred to as AVs) play a crucial role in modern society by preventing and detecting malware. Ideally, AVs should precisely detect known malware, rapidly respond to newly created malware, and recognize new features of variants. However, in reality, their effectiveness falls short. Modern AV engines not only perform inadequately against newly created malware or attack vectors~\cite{malwarebytes}, but also face challenges in detecting variants of known malware.

Prior studies~\cite{droidchameleon2013,asiaccs2016mystique,mystiques2017,empirical2018icse} have made pioneering efforts to evaluate AV engines by evolving Android malware, largely with obfuscation. Although they can reflect the AVs' resistance to obfuscated malware, the results suffer from two limitations. First, apart from the transformed code, obfuscation can cause inconspicuous changes to an app, such as the destruction of app certificates and hash codes. Therefore, it cannot necessarily imply that obfuscation is the main reason for evasion. Second, modern AVs are usually complex systems that combine multiple detection techniques~\cite{symantec,google} and harvest features from various modules in an app. This complexity cannot guarantee that obfuscation has precisely destroyed AVs' key features and thereby caused evasion. There are many existing evasion techniques that can effectively bypass detection, such as sandbox evasion~\cite{yokoyama2016raid,miramirkhani2017sp} and adversarial attacks~\cite{adv2020sp}. Therefore, it is intriguing and significant to understand how an AV engine detects malware and its robustness to malware variants. Different from obfuscation-based approaches~\cite{droidchameleon2013,empirical2018icse}, our study aims to identify what clues from an app, i.e., indicators of compromise (IoC), are likely collected by AV engines and how they are used. However, given an app, AVs usually work as black boxes and only return limited information (e.g., malware name). We cannot even ensure whether the detection is correct, let alone speculate how the result is produced.

In this study, we propose a meticulous evolution of Android malware and employ an interaction model to identify the IoCs used by AV engines. Since there are thousands of AVs circulating in the market, deploying them locally becomes prohibitively expensive and inefficient. We instead resort to the online scanning service, \virustotal, as a substitute, which runs a number of off-the-shelf AVs at the backend and has been proven to rival, even surpass, desktop ones~\cite{virustotal2020usenix}. To accomplish this goal, we collect 23,184 Android apps, including malware, grayware, benign apps, and white-listed apps as the experimental subjects (Section~\ref{sec:approach:data}). We select six app transformation techniques—unsigning, re-signing, pruning, fusing, packing, and dynamic-loading—to change the IoCs in an app, including fingerprints, certificates, code, and dynamic features (see Section~\ref{sec:approach:transform}). By collecting the scanning reports from \virustotal for these apps and their transformations, we obtain a massive number (971K) of detection reports from over 70 antivirus engines. Then we employ data-driven approaches to quantify and analyze these differences among apps and further infer the detection mechanism running inside AVs. Finally, substantial analysis results, as well as 6 insightful findings, are presented in Section~\ref{sec:eval}, shedding light on the incautious use of online scanning services in recent research and highlighting a number of issues during malware detection that could be improved.

After the analysis, we have highlighted the following findings. 
It is measured that \emph{unsigning} and \emph{re-signing} (with an AOSP key and self-signed key) can decrease the maliciousness of apps by 31.4\%, 25.4\%, and 24.2\%, respectively. However, we have found evidence from the data that these drops are largely attributed to the availability of the code of blacklisted apps and the validation of signing certificates during malware detection (see Section~\ref{sec:eval:signature}).
App code, native code, and XML files exhibit different importance when AVs get features as malware evidence, and the pruning to these modules averagely reduces the maliciousness by 53.5\%, 25.2\%, and 23.5\%, respectively. With code modularization in Section~\ref{sec:eval:static:prune}, we have successfully identified which parts of the code contribute the most to maliciousness (see Section~\ref{sec:eval:static:prune}).
Some AVs are still able to correctly recognize packed apps with consideration of unencrypted files located in the resources folders (e.g., \texttt{assets}, \texttt{libs}, and \texttt{res}). The majority of AVs can deal with compressed payloads, but none of them are able to detect malicious payloads which are split into multiple files (see Section~\ref{sec:eval:static:pack}).
Only one online AV engine can successfully identify the apps that dynamically load malicious payload during the 7-day tracking. It reveals the incapability of dynamic analysis (see Section~\ref{sec:eval:dynamic}).
Many sandboxes deployed in \virustotal perform a security check on URLs contained in apps, but few of them dynamically execute apps as observed. By inserting logic bombs, we can evaluate the capability of dynamically analyzing apps in Section~\ref{sec:eval:dynamic}.

\noindent\textbf{Contributions.} We have made the following contributions:
\begin{itemize}[leftmargin=*]
	\item \textbf{Massive number of transformed Android apps and detection results.} We develop six transformation techniques, and generate over 179K transformed apps in total from 23,184 Android apps of multiple sources. Through querying online scanning services, we obtain 971K security reports in a time frame of 30 days. We will open-source our transformation code, security reports from \virustotal, and non-private analysis results to the public in future\footnote{All transformation code and transformed apps can be accessed from https://github.com/impillar/AVScale. Due to the huge volume of apps, we only provide the hash codes for these apps, and the reports can be downloaded from \virustotal.}. 
	\item \textbf{Analysis and explanation of the effects of app transformations.} We conduct a comprehensive and extensive analysis based on the collected data and reports to demystify AVs in accordance with signature-based, static analysis-based detection, and dynamic analysis. We further make the attempt to explain these results and highlight six findings that unveil previously unknown phenomena during malware detection. 
\end{itemize}

\section{Background}\label{sec:background}

\subsection{Commercial Antivirus Software}\label{sec:background:cav}
Antivirus software is a computer program used to prevent, detect and remove malware. Towards the ever-increasing malware, it employs multiple techniques, such as signature-based, heuristic approaches, real-time detection, to confront the threats from malware. The complexity and diversity of malware drive the advancements of modern AV software, and born many security vendors. 
Most of AVs work as blackbox systems, making it unclear how a file to test goes though AVs and returns with a detection label. According to~\cite{koret2015antivirus}, several detection mechanisms are probably encapsulated within an AV engine, as elaborated below.
\noindent\textbf{Malware signature.} Signatures are a simple yet effective feature for recognizing known malware~\cite{signature2009raid,signature2017asiaccs}.
Intuitively, all objects have their unique signatures. So one app is malware theoretically, if its signature is in the malware corpus. 
Consequently, the adoption of signatures can significantly enhance accuracy while reducing false positive rates in malware detection.  
Signature can be created in varying manners~\cite{koret2015antivirus}, for example, cryptographical hashes (\eg, MD5, SHA1 and SHA256), specific strings.  
Cryptographical hash is a digest that can uniquely represent one object; even a minor alteration to an app results in a drastically different hash. Strings like IoCs can be a specific piece of code.
According to our investigation, it is widely acknowledged and incorporated by many security vendors~\cite{symantec,anva}. 
Moreover, fuzzy logic-based signatures (\eg, ssdeep~\cite{ssdeep}) have attracted many security vendors where chunks of data, rather than the whole file, are computed for hashes. This can mitigate the weakness of the-whole-file signature against a slight change to malware. 


\vspace{3pt}
\noindent\textbf{Static analysis}. This is a detection method that does not actually execute files. By gathering features and information from code, it can thereby determine the maliciousness of one file to test. Additionally, there are a series of determination methods like pattern matching, machine learning~\cite{continuous-detection23,botacin25cs}, and formal reasoning. Usually, static analysis based malware detection is an effective approach that can access all actionable code and semantic logics inside. From the simple pattern matching to complicated constraint solving, nearly all AVs have employed static analysis for detection.

\vspace{3pt}
\noindent\textbf{Dynamic analysis}. Different from static analysis, it needs to execute the code for harvesting the runtime information. To achieve this target, an AV should be equipped with a sandbox which physically installs and executes the code. Usually, instrumentation is required for monitoring system status and code behaviors. This method has lower false positives compared to static analysis, but suffers from test coverage and logic bombs~\cite{triggerscope2016sp}.




\subsection{Online Scanning Services}
Intuitively, one single AV engine is prone to be subverted by targeted attacks, so that more and more users start to assemble multiple AV engines to provide security protection. 

\vspace{3pt}
\noindent\textbf{\virustotal}. It is a public website providing multiple security services, for example, file scanning, URL scanning, and data correlation analysis. To date, it has integrated 76 antivirus engines from 72 security vendors~\cite{virustotal}, providing a real-time scanning service (note that security vendor Avast has two products--Avast and Avast Mobile Security equipped in \virustotal, and so do K7 Computing, Symantec Corporation, and Trend Micro). Additionally, a number of sandboxes are deployed to perform behavioral analysis on the programs to test.  
In this study, we find 97 antivirus engines from collected 971K security reports although some of them appear with limited times. We further filter these AVs by removing those that return over 90\% ``type-unsupported'' results and obtain 66 ones. 

For an Android app, \virustotal offers a number of miscellaneous APIs for developers to fetch security analysis reports, scan suspicious files, retrieve the detailed information of app and so on. 
The analysis results by antivirus engines are presented with a 6-tuple ``\emph{(category, engine\_name, engine\_update, engine\_version, method, result)}''. In particular, \emph{category} denotes the marked flag by antivirus which could be ``confirmed-timeout'', ``failure'', ``harmless'', ``malicious'', ``suspicious'', ``timeout'', ``type-unsupported'', ``undetected''. If the app is malware, the attribute ``result'' will present its most likely name, otherwise is null.

\section{Preliminary}\label{sec:model}

\begin{figure*}
	\centering
	\includegraphics[width=1\textwidth]{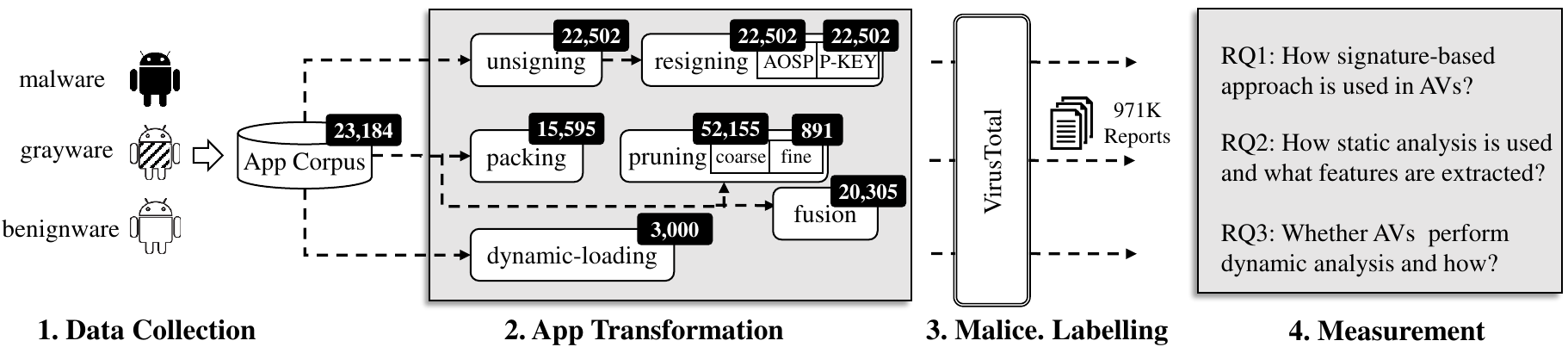}
    \caption{System overview of our approach. It contains six app transformations and the numbers in black count generated apps.}\label{fig:overview}
	\vspace{-3mm}
\end{figure*}

For users, antivirus engines operate as a black box, accessible only through limited interfaces. The behavior and reactions of antivirus engines to different apps remain unknown and mysterious. When scanning an app, users can only determine whether it is malicious and its malware type as reported by the antivirus engine.
Consider a malware app $A_1$. When we create a transformed app $A_2$ from $A_1$, the antivirus engine may yield different results for $A_1$ and $A_2$. Let $\Delta_{a}$ represent the difference between the apps, and $\Delta_{r}$ denote the difference between the corresponding detection results. We hypothesize that the cause of $\Delta_{r}$ is attributable to $\Delta_{a}$, i.e., $\Delta_{a}$ $\implies$ $\Delta_{r}$. This assumption is also employed by prior studies~\cite{malgene2015ccs,asiaccs2016mystique,empirical2018icse} to evaluate the effectiveness of malware evasion techniques.

\noindent\textbf{Interaction Model}. Without loss of generality, we define the interaction model with blackbox antivirus engines as follows. Let $\Sigma$ be the set of apps to test, $\mathcal{F}$ is a blackbox antivirus engine, and $\mathcal{F}: \Sigma~\rightarrow~(\mathcal{B}, \mathcal{L})$, where $\mathcal{B}$ is a \textit{boolean} value and $\mathcal{L}$ is the output labels of tested apps. Note that if $\mathcal{B} = false$, $\mathcal{L} = null$ and the app is recognized as benign.


Moreover, we assume $Tr$ as the transformations between apps with traceable changes, so $\Sigma \times Tr \rightarrow \Sigma$. 
Given an input app $u$ ($u\in\Sigma$), the output by the antivirus engine can be denoted as $(\mathcal{B}_u, \mathcal{L}_u)$. 
The transformed app $v$ can be created by applying $v = Tr(u)$, and we subsequently get its output label $(\mathcal{B}_v, \mathcal{L}_v)$. 
Let $\Delta_a$ be the difference between apps, \ie, $\Delta_a = v - u$, $\Delta_r$ be the difference between output labels within an transformation, \ie, $\Delta_r = (\mathcal{B}_u - \mathcal{B}_v, \mathcal{L}_u - \mathcal{L}_v)$. 
$\Delta_r$ accounts for the following two categories.
\begin{itemize}[leftmargin=*]
	\item \textbf{Maliciousness flip - $\mathcal{B}$}. The decision made by an antivirus engine shifts from benign to malicious or conversely. This can be recognized with the transitions from ``undetected'' to ``malicious'' in \virustotal reports, which is noted as a \emph{negative flip}. A \emph{positive flip} is a transition from ``malicious'' to ``undetected''.
	\item \textbf{Label change - $\mathcal{L}$}. For two malicious samples, an antivirus engine may report with different malware families. For example, the attribute ``result'' in \virustotal may shift from ``FakeInst'' to ``SmsReg''.
\end{itemize}

In particular, maliciousness flip reflects that the transformation has cracked the indicator employed by AVs for detection, and label change implies that the modifications destroy the original indicator but produce another one with transformation. 

\noindent\textbf{Our Goals.} 
An antivirus engine may employ three techniques for malware detection including signature-based, static and dynamic analysis.
For each detection technique, we intend to identify the IoCs used by AVs and how these IoCs are used for recognizing malware. To be specific, we attempt to explore:

\begin{enumerate}[label=\textbf{RQ\arabic*.},leftmargin=*]
	\item How do these antivirus engines use signature for detection and are there any weaknesses (see Section~\ref{sec:eval:signature})?
	\item How do these antivirus engines use static-based detection and what features matter (see Section~\ref{sec:eval:static})?
	\item Is dynamic analysis widely used in these engines and how (see Section~\ref{sec:eval:dynamic})?
\end{enumerate}



\begin{table}
	\centering
	\scriptsize
	\caption{Sources of Android apps and the use in experiments}
	\label{tab:data}
	\begin{tabular}{cc|cl} \toprule
		\multicolumn{2}{c|}{\textbf{Dataset}} & \textbf{\# Size} & \textbf{Experiment} \\ \midrule
		\multirow{3}{*}{Malware} & \genome~\cite{genome2012} & 1,235 & Section~\ref{sec:eval:signature}, \ref{sec:eval:static}, \ref{sec:eval:dynamic}, \ref{sec:eval:report} \\ 
		& Drebin~\cite{drebin} & 5,555 & Section~\ref{sec:eval:signature}, \ref{sec:eval:static}, \ref{sec:eval:dynamic}, \ref{sec:eval:report} \\ 
		& \virusshare~\cite{virusshare} & 6,017 & Section~\ref{sec:eval:signature}, \ref{sec:eval:report} \\ \midrule
		Grayware & AMD~\cite{amd2017dimva} & 2,659 & Section~\ref{sec:eval:signature}, \ref{sec:eval:static}, \ref{sec:eval:report} \\ \midrule
		\multicolumn{2}{c|}{ANVA apps} & 996 & Section~\ref{sec:eval:report} \\ \midrule
		\multicolumn{2}{c|}{Wild apps} & 7,074 & Section~\ref{sec:eval:signature}, \ref{sec:eval:static}, \ref{sec:eval:report} \\ \midrule
		\multicolumn{2}{c|}{\textbf{Total}} & 23,184 &  \\ \bottomrule
	\end{tabular}
 \vspace{-3mm}
\end{table}
\section{The Approach}\label{sec:approach}

Figure~\ref{fig:overview} shows the overview of our approach. 
It proceeds with four phases: data collection, app transformation, maliciousness labelling and measurement. Specifically, we prepare a set of Android apps (totaling 23,184) for our study, including malware, grayware, ANVA apps and wild apps. 
These apps are transformed by six techniques, and uploaded into \textsc{VirusTotal} for labelling. 
Last, we undertake the analysis of security reports to answer the questions in Section~\ref{sec:model}.

\subsection{Data Collection}\label{sec:approach:data}
To ensure diversity, we gather four types of Android apps: malware, grayware, ANVA apps, and benign apps. These categories differ in the number of malicious indications flagged by AVs. Typically, malware samples exhibit more malicious indications from AVs compared to grayware and benign apps.


\noindent\textbf{Malware.} It encompasses Android apps designed with malicious intent, aiming to compromise Android systems and users for various purposes such as privacy theft, privilege escalation, and unauthorized premium service charges. 
In this study, we select 6,017 samples from \virusshare~\cite{virusshare} in the past five years, 1,235 samples from \genome~\cite{genome2012}, 5,555 samples from \drebin~\cite{drebin}. 

\noindent\textbf{Grayware.} Grayware, while not as harmful as malware, is still problematic as it often consumes device resources excessively, displays annoying advertisements, or solicits users' private information frequently. We randomly select 2,659 grayware samples from the AMD dataset~\cite{amd2017dimva}, covering eight known families including Airpush, Andup, Dowgin, Kuguo, Kyview, Minimob, Utchi, and Youmi.

\noindent\textbf{ANVA apps.} Anti Network-Virus Alliance is an authoritative organization and publishes a number of whitelisted apps every year~\cite{anva}. These apps have been vetted by 11 renowned security vendors. We collect 996 whitelisted apps from them to evaluate how AVs react to their transformations. 

\noindent\textbf{Benign apps.} To provide a comprehensive evaluation of antivirus engines, we additionally crawl 7,074 Android apps from the wild, including the official app store Google Play, and alternative ones such as Apkpure~\cite{apkpure}. None of these apps are flagged as malware by any AV engines in \virustotal.
 
Table~\ref{tab:data} shows the statistics of Android apps and their utilization in our experiments. Our objective is to explore how AV engines respond to transformed apps, necessitating that AVs can ideally identify their original versions. Therefore, we select \drebin, \genome, AMD, and \virusshare datasets, all of which are publicly available.
Additionally, we found multiple duplicates across these datasets. There are 909 duplicated apps residing in both \drebin and \genome datasets, 224 apps in \virusshare and Grayware, 6 apps in both \drebin and \virusshare, 1 app in \drebin, \genome and \virusshare, and one app in ANVA is also found in \virusshare. After removing these duplications, we obtain 23,184 unique apps.

\definecolor{Gray}{gray}{0.9}

\subsection{App Transformation}\label{sec:approach:transform}

We present six app transformation techniques in this section, and briefly introduce the implementation for them. 

\noindent\textbf{Design Principle.} Prior research on malware evasion and evaluation employs many program transformation techniques like obfuscation~\cite{droidchameleon2013}, refactoring~\cite{refactor2015ase}, shrinking~\cite{jcs2019panguard}, optimization~\cite{jcs2019panguard}, and packing~\cite{packer2018ndss}. Many of them are based on an assumption that AVs must gather features from code for detection, so transforming code will definitely influence detection results. 
However, this assumption may not apply to all AV engines~\cite{droidevolver2019}. It remains unclear what features are collected by blackbox AVs and how they are used. Therefore, we dissect one APK file and identify five components which can be the source for feature harvest: file signature, DEX file, XML files, asset files and certificate. 
Subsequently, we design six transformation techniques which can modify these features with fine control. 
Although these transformations are not all in the real world, they are effective in finding the answers of the research questions in Section~\ref{sec:model}.

\subsubsection{Unsigning} Android apps are required to be digitally signed with a private certificate before shipment to ensure their integrity. Apps without a valid certificate cannot be installed or executed on a device by default. Therefore, a malware app with a corrupted or missing certificate poses no risks. To determine whether antivirus engines verify the integrity of apps under test, we propose an \emph{unsigning} transformation to remove the certificate of the app. We implement this transformation by disassembling a signed app and then assembling it into an unsigned app using \apktool. The unsigned app will have a different hash code (e.g., SHA256 and MD5) compared to the original. Consequently, if an antivirus engine relies solely on the app's digest without considering other features, unsigned apps are likely to evade detection.

\subsubsection{Re-signing} Th certificate of Android apps has been used to identify repackaged apps~\cite{repackage}. That motivates us to explore whether blackbox AVs have used app certificates for detection. Here we use two certificates for app signing. One is a self-signed certificate, which must be the first sight by AVs. 
The other is certificates from Android Open Source Platform (AOSP), which are likely seen for many times. It is observed that many developers use AOSP certificates to sign their own apps. Re-signing can be achieved with the tools \emph{jarsigner} and \emph{apksigner}. 
Additionally, if too many apps are signed with the same certificate and sent to AV engines, it may raise the awareness of engines and produce unpredictable results. 
For eliminating the cross effect among transformed apps, we create different certificates for different tasks. 

\subsubsection{Pruning}
There are several building block files for Android app, including Java/Kotlin code (compiled into dex files), native code, XML file. 
To evaluate how AVs leverage these files for determining maliciousness, we propose to prune an app by removing parts of contained files. 
We develop three strategies for pruning an app: 1) eliminating partial or all actionable code in the app. Specifically, we remove all statements in methods while leaving the method declarations unaltered. 
For the methods with a \texttt{return} value, we make them return \texttt{Null} all the time. 
With this pruning, all malicious code in Java  can be excluded; 2) eliminating native code that is oftentimes stored in the folder ``\texttt{lib}'' and  ``\texttt{asset}''. 
It is easily implemented by deleting all the native code in an app, which cannot hinder successful compilation. Therefore, the malware whose malicious code reside in native cease to effect; 3) removing all configuration information from the AndroidManifest.xml file that do not influence app packaging, such as \texttt{\small <permission>}, \texttt{\small <permission-group>}, \texttt{\small <permission-tree>}, \texttt{\small <users-permission>}, \texttt{\small <uses-configuration>},  \\ \texttt{\small <uses-feature>}, \texttt{\small <users-library>} and \texttt{\small <uses-sdk>}.

\subsubsection{Fusion} Apart from pruning that reduces functions from apps, we fuse two apps to combine the functions of different apps.  
Especially when two malware are merged together, it is intriguing to explore how AVs react to more complex or compound malware.
We develop a transformation technique to automatically merge two apps by specifying the host app and the payload app. 
To resolve the conflicts during fusion, we particularly process the following files:
\begin{itemize}[leftmargin=*]
    \item AndroidManifest.xml: We compute the union set for specific elements such as ``\texttt{\small <uses-permission>}'', ``\texttt{\small <permission-group>}'', \\ ``\texttt{\small <uses-feature>}'', ``\texttt{\small <uses-library>}''. For the components of the fused apps, we pick one of them as the host app, and keep all its component while removing the \texttt{intent-filter} of the main activity of the other. 
    \item Resources files: Most of resource files reside in the folder \texttt{res}, such as GUI layouts (\eg, \texttt{layout/*.xml}), drawable images (\eg, \texttt{drawable/*.xml}), string used in app (\eg, \texttt{values/string.xml}), and id files (\eg, \texttt{values/public.xml}). 
    Many of these files can be glued by computing their file union set. We merge the content of files if two apps possess the same files.
    \item App code: We merge code into one single folder. If there exists another file with the same name, we only remain one of them. Although it may lead unexpected compilation errors, we avoid to resolve this conflict but only use the fused apps with successful compilation in our experiments.
\end{itemize}

\subsubsection{Packing} Packing is a protective measure for software against reverse engineering and arbitrarily code tampering. However, it is also appealing to malware authors for concealing their malicious code in an app~\cite{packer2018ndss}. 
Generally, Android packing first encrypts the original DEX files and then creates a proxy class inherited from ``\texttt{android.app.Application}'', ensuring it is the first class to execute. 
The proxy class will decrypt the primary DEX files, and execute them dynamically. 
This process prevents static analysis from extracting the original code and detecting malware.
To uncover how antivirus engines react to these packed benign or malicious apps, and whether they have the ability to detect malicious code inside the packer, we propose to automatically pack a number of benign and malware samples with \cite{bangle-code}, and feed them to AVs.

\subsubsection{Dynamic-loading} For the seek of flexibility, Android supports components to be dynamically created and executed with reflection. It can be used for hot patching~\cite{hotpatch} and automatic app generation~\cite{generator}. However, it can be also employed by malware authors for dynamically loading their malice and then launching attacks~\cite{execute2014ndss,mystiques2017}. 
Since all malicious code is outside of the app space, AVs that rely solely on static analysis cannot capture its malicious code but may be aware with the dynamic loading behaviors. 
Without dynamic analysis, AVs cannot recognize the exact malicious behaviors in the payload.
In this study, we create a proxy app that only downloads malicious payload from the cloud, and aim to identity whether dynamic analysis is conducted in AVs.

\subsection{Maliciousness Labelling}

In this study, we rely on \virustotal as the oracle to determine the maliciousness of apps. 
Each \virustotal report provides over 70 data points, where one data point consists of six properties:\emph{category}, denoting the type of detection results, \emph{engine\_name}, means the name of engine, \emph{engine\_update} is when the engine is updated, \emph{engine\_version} is the identical version, \emph{method} reveals how antivirus engines detect, and \emph{result} indicates the label of malware. To fetch plenty of detection reports for analysis, we collect two types of security reports as follows:

\begin{itemize}[leftmargin=*]
	\item \emph{First snapshot}. Online scanning services respond to authorized requests with detection results by AVs. 
    We collect the staged reports for already-analyzed apps, otherwise, upload apps for scanning. 
	\item \emph{Reanalysis snapshot}. It is observed that \virustotal constantly upgrades its hosted AVs and re-test the apps stored on the server. Due to the updates in the malware blacklist, enhancement of AVs' detection capabilities, and so on, the labels of app maliciousness may change over time~\cite{virustotal2020usenix}.
	Therefore, we request a forceful re-analysis of these apps to get the latest detection results. 
\end{itemize}

After fetching a security report from \virustotal, we can determine the maliciousness of the app. 
Without loss of generality, we define the maliciousness of an app by the number of AVs that recognize them as malware. We assume that $AV$ is the set which supports the detection of our apps. Given an app $u$, its maliciousness can be represented as $M(u)~=~|AV_{i}|~where~\mathcal{L}_{i}(u) \in malware$.






\section{Measurement}\label{sec:eval}

%




In this section, we present the analysis corresponding to the research questions in Section~\ref{sec:model}. 


\subsection{Signature-based Detection (RQ1)}\label{sec:eval:signature}
It is unclear how antivirus engines adopt signatures for malware detection. In this section, we conduct the following experiments, and evaluate their signature-based detection.

\begin{figure*}[t]
	\centering
	\begin{subfigure}[t]{0.32\textwidth}
		\centering
		{\includegraphics[width=1.0\textwidth]{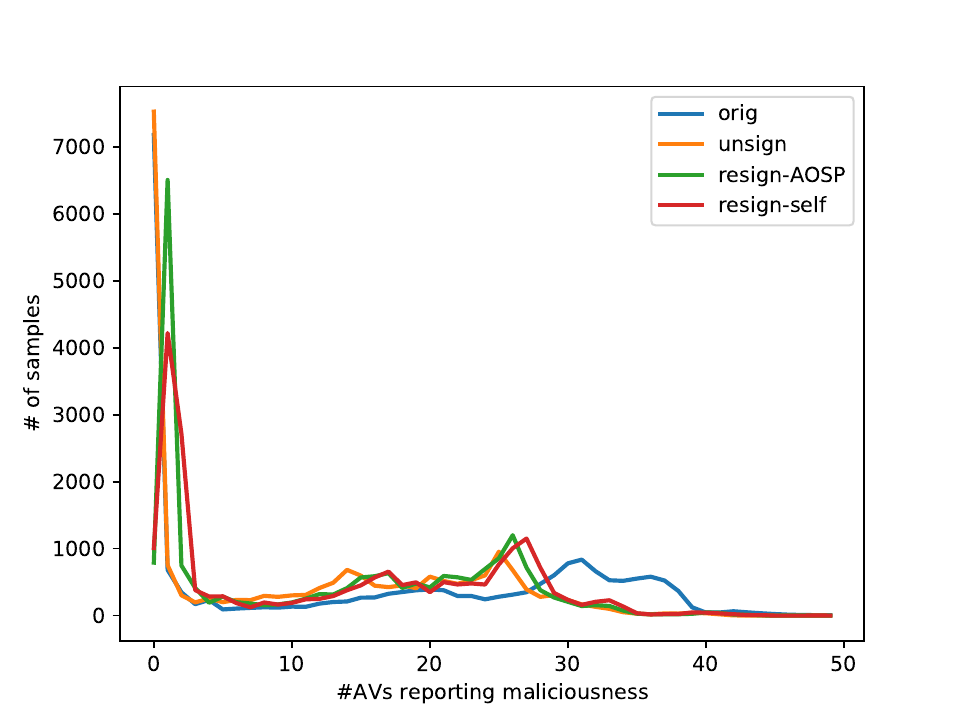}}
		\caption{Distribution of apps vs. \# detected AVs} 
		\label{fig:all-sign}
	\end{subfigure}
	\hfill
	\begin{subfigure}[t]{0.32\textwidth}
		\centering
		{\includegraphics[width=1.0\textwidth]{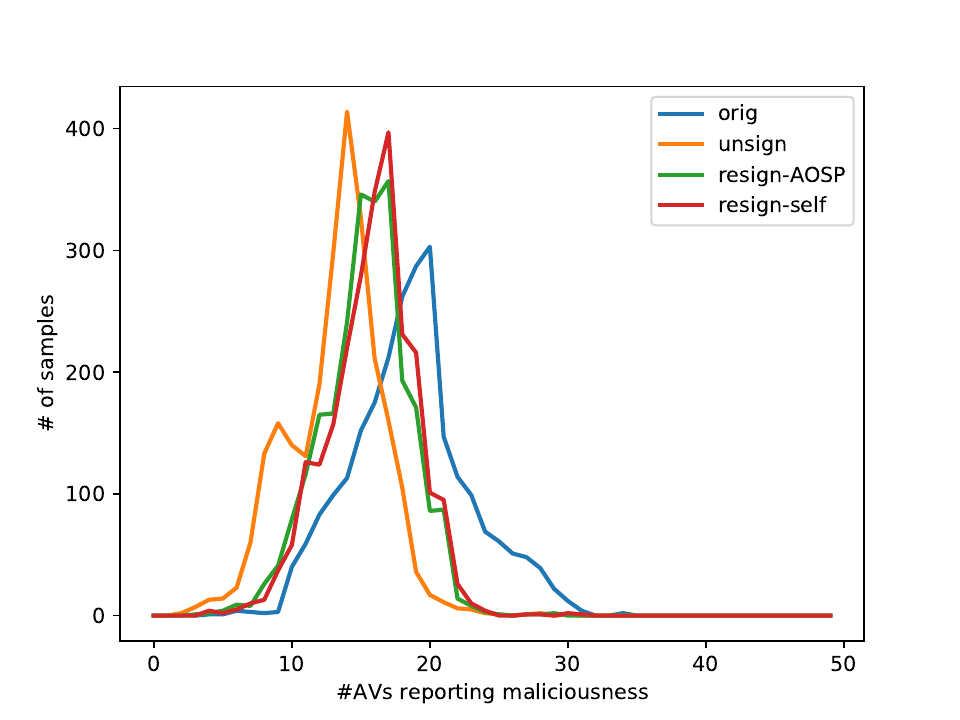}}
		\caption{Dist. of grayware vs. \# detected AVs} 
		\label{fig:grayware-sign}
	\end{subfigure}
	\hfill
	\begin{subfigure}[t]{0.32\textwidth}
		\centering
		{\includegraphics[width=1.0\textwidth]{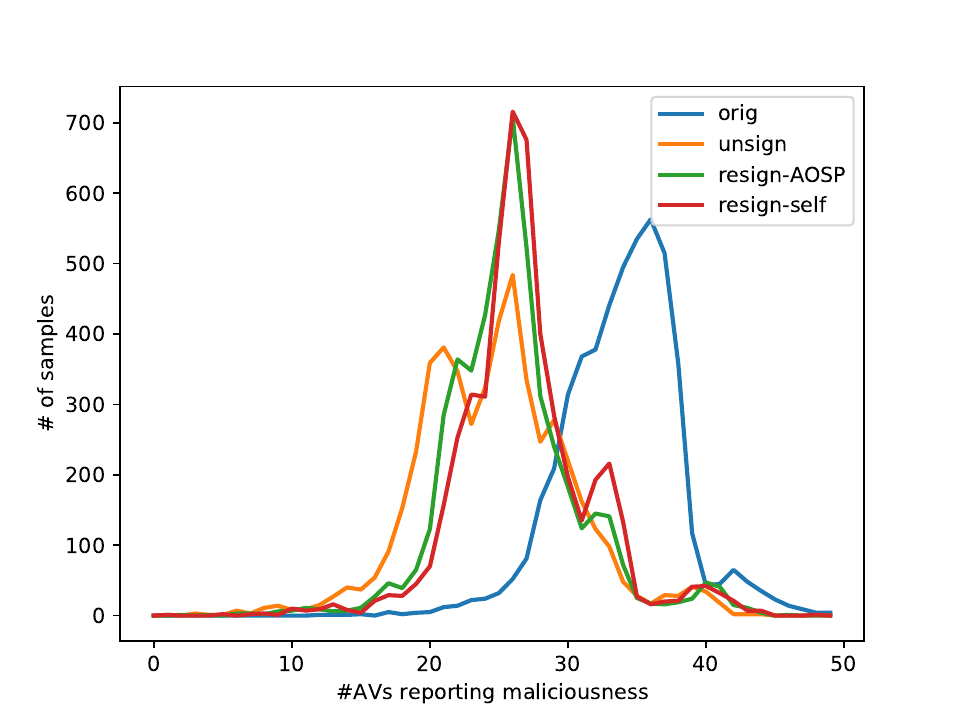}}
		\caption{Dist. of malware vs. \# detected AVs}
		\label{fig:malware-sign}
	\end{subfigure}%
	\caption{Distribution for apps in terms of the number of detected AVs. The curves in blue represent the original apps, orange denotes the unsigned apps and green curves shows the distribution for the re-signed apps.}\label{fig:unsign}
 \vspace{-3mm}
\end{figure*}

\noindent\textbf{Experiment Design}. In this experiment, we try to crack apps' certificate as well as their hash code and investigate how AVs react afterwards. We employ three strategies by: 
\begin{enumerate}[leftmargin=*,label=\textbf{S\arabic*.}]
	\item removing the signing certificate for each app, from which we obtain 22,502 unsigned apps in total; 
	\item re-signing all the apps of S1 with a publicly-available certificate on Android Open Source Platform (AOSP); 
	\item creating a self-signed certificate and re-signing all S1 apps.
\end{enumerate}


These strategies are devised with two considerations. On one hand, app unsigning and re-signing are the two minimal changes that can be performed on Android apps while preserving app integrity. 
Any trials to alter code or files can surely effect app signing or destroy its integrity. 
On the other hand, null, self-signed and public certificates serve as control variables, and we can infer whether and how these antivirus engines take advantage of them. 
We present Figure~\ref{fig:unsign} for a quick overview of the change of app maliciousness. 
More specifically, a data point ($x$, $y$) in Figure~\ref{fig:all-sign} means there are $y$ apps of which the maliciousness is $x$. We plot four curves for the original apps, and their transformations respectively. 

\vspace{3pt}
\noindent\textbf{Result Analysis.} 
From Figure~\ref{fig:unsign}, we can see a remarkable left shift for transformed apps, \ie, unsigned, re-signed apps exhibit less maliciousness compared to the original. The three strategies can reduce the degrees of maliciousness by 28.2\%, 19.4\% and 18.5\%, respectively. Exceptionally, there is an abrupt increase and decline with S2, and S3-transformed apps with $x$ being zero. Among 7179 apps with $M=0$, 93.9\% of their transformed apps start to exhibit more or less maliciousness. 
Overall, the positive flip rates (PFR) across all AVs are (31.4\%, 25.4\%, 24.2\%) for the three strategies in the first snapshot, but decrease to (15.4\%, 12.8\%, 23.3\%), and $PFR = \frac{\#~ positive ~flips}{\# ~malware~labels}$. 
The transformations of \emph{unsigning} and \emph{re-signing} with AOSP keys significantly impact maliciousness in subsequent snapshots, while \emph{re-signing} with self-signed keys has less effect. This underscores the importance of considering the repackaging effect in malware transformation.

Table~\ref{tab:shift} displays the top 5 AVs with the highest and lowest PFR. A higher PFR indicates greater susceptibility to repackaging, implying heavy reliance on app hash codes for detection. Variance in PFR among AVs suggests differing detection methods:

\begin{itemize}[leftmargin=*]
	\item \textbf{High S1-, S2-, S3-PFRs.} We cluster malware samples recognized by VBA, Zillya, TrendMicro-HouseCall, TrendMicro into multiple families, and check whether these families have similar PFRs. We find that VBA have a superior capability of recognizing families ``BaseBridge'' (98.7\%), Zillya is sensitive with ``DroidKungFu'' (66.2\%), TrendMicro-HouseCall achieves 85.1\% recall with ``DroidKungFu'' and TrendMicro gets 80.0\%. These exceptions reveal that these AVs have used other methods to recognize malware when the hash code does not match. However, AVs possess a biased set of malware samples from which they harvest signature for detection. Obviously, ``DroidKungFu'' and ``BaseBridge'' are the two families with ample features.
	\item \textbf{S1-PFR $\gg$ S2-PFR $\approx$ S3-PFR.} Zoner (90.4\%, 24.2\%, 26.3\%) is the only instance exhibiting this characteristics. It can be implied that Zoner first verify the integrity of the scanning target since an app without certificates cannot be installed and pose risks.  
	\item \textbf{S3-PFR $\gg$ S2-PFR.} For example, Alibaba (S2=2.1\%, S3=97.7\%), McAfee-GW-Edition (S2=4.2\%, S3=74.8\%), MAX (S2=25\%, S3=94\%), Symantec (S2=10.7\%, S3=59.0\%), and Tencent (S2=9.6\%, S3=50.0\%). It shows that these AVs have a definite recognition of app certificate and take it as an important, or even dominating clue for malware. Surprisingly, a malware sample can easily evade these AVs' detection by simply using a new signing certificate. Particularly, MAX, a machine learning-based engine, obviously takes signing certificate as a dominating feature in classification.
\end{itemize}


\begin{table}
	\caption{AVs with the highest/lowest positive flip rates for S1, S2 and S3 apps. We only consider the AVs that recognize at least 2,000 malware samples.}\label{tab:shift}
	\resizebox{\linewidth}{!}{
	\begin{tabular}{cccccc} \toprule
		\textbf{AV} & \textbf{\# Apps ($>$2K)}& \textbf{S1 (\%)}& \textbf{S2 (\%)} & \textbf{S3 (\%)}  & \textbf{Ave. (\%)} \\ \midrule
		VBA32 & 2,712 & 78.6 & 78.6 & 78.6 &	78.6 \\ 
		Zillya & 2,399 & 69.0 & 80.6 & 79.0 & 76.2 \\ 
		TrendMicro-HouseCall & 3,044 & 73.3 & 74.5 & 73.2 & 73.7	 \\ 
		TrendMicro &  3,189 & 72.6 & 73.8 & 72.9 & 73.1 \\ 
		Jiangmin & 7,661 & 64.5 & 64.5 &	65.0 & 64.6	 \\ \midrule
		Kaspersky &	9,341 &	1.0 & 1.9 & 1.4 & 1.4 \\		
		Avira &	12,433	&	1.0 & 0.8 &	2.5 & 1.4 \\
		ZoneAlarm	& 9,371	& 1.0 &	1.6 &1.4 &1.3 \\		
		Trustlook &	12,236	&	2.1 & 0.7 &	0.4 & 1.1 \\
		ESET-NOD32 & 12,959	& 0.3 &	0.2 & 0.3 &	0.2 \\
		\bottomrule
	\end{tabular}
}
\vspace{-5mm}
\end{table}


A negative flip occurs when a benign label becomes malicious. Most AVs exhibit a negative flip rate (NFR) below 10\%, with Trustlook being the exception (7.6\%, 89.1\%, 86.9\%). In S2, 89.9\% of negative flips show the label ``Android.PUA.DebugKey,'' indicating Trustlook's focus on certificate verification before classification.
Of the 7,129 apps in our corpus rated as non-malicious, 20.1\% are classified as malware by at least one AV after the three transformations. Trustlook is responsible for 65.3\% of negative flips, K7GW 27.7\%, while Kaspersky and ZoneAlarm contribute 2.4\% each.


\begin{figure*}[t]
	\centering
	\begin{subfigure}[t]{0.3\textwidth}
		\centering
		{\includegraphics[width=1.0\textwidth]{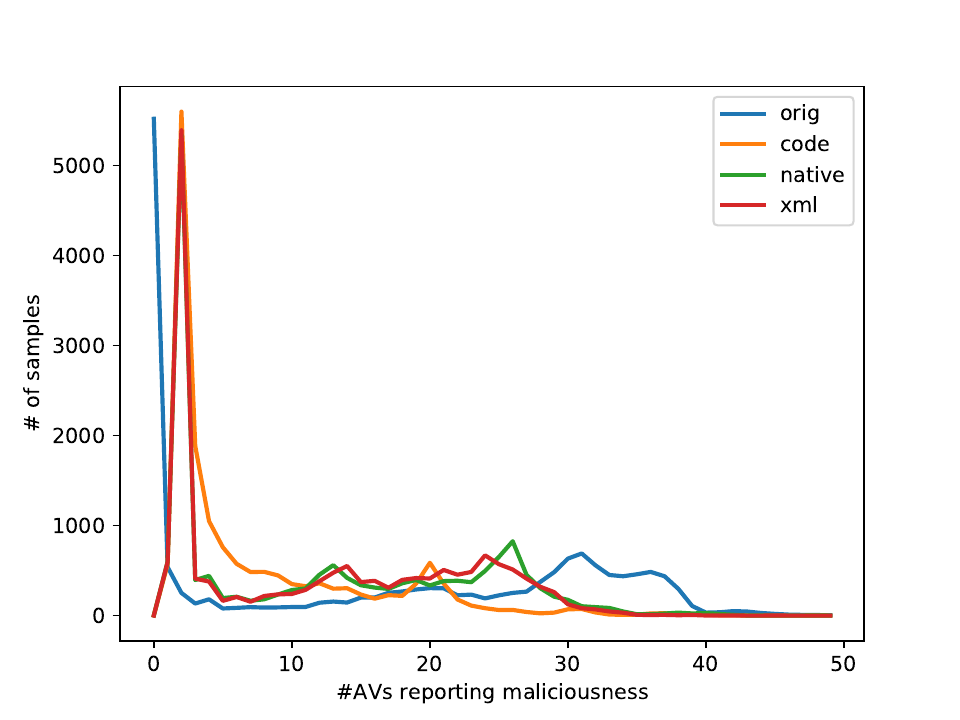}}
		\caption{Distribution of apps vs. \# detected AVs} 
		\label{fig:all-prune}
	\end{subfigure}
	\hfill
	\begin{subfigure}[t]{0.3\textwidth}
		\centering
		{\includegraphics[width=1.0\textwidth]{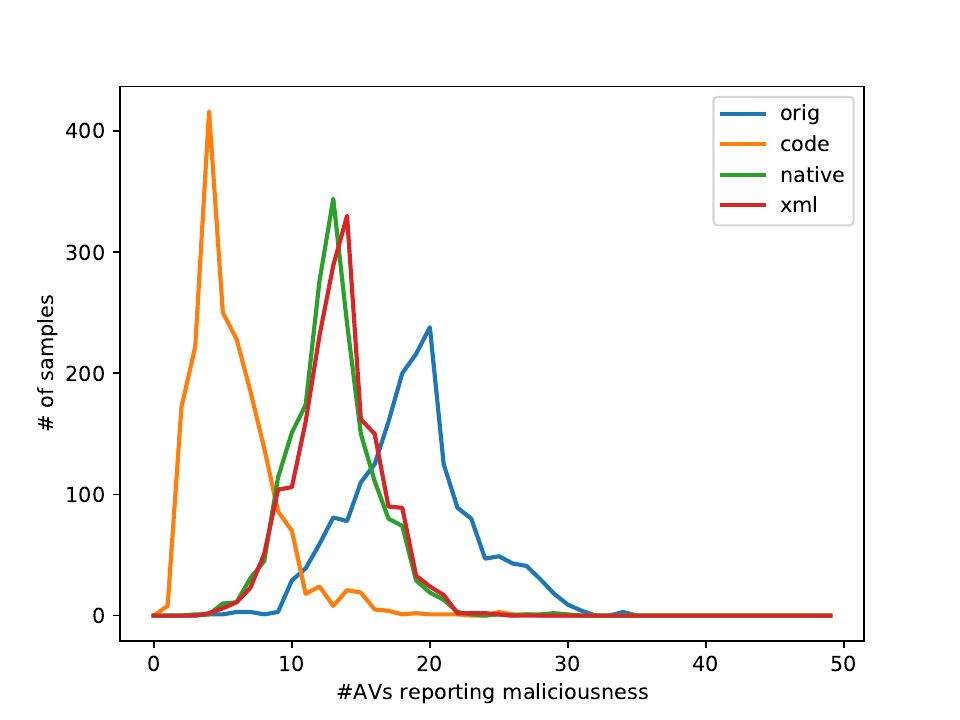}}
		\caption{Dist. of grayware vs. \# detected AVs} 
		\label{fig:grayware-prune}
	\end{subfigure}
	\hfill
	\begin{subfigure}[t]{0.3\textwidth}
		\centering
		{\includegraphics[width=1.0\textwidth]{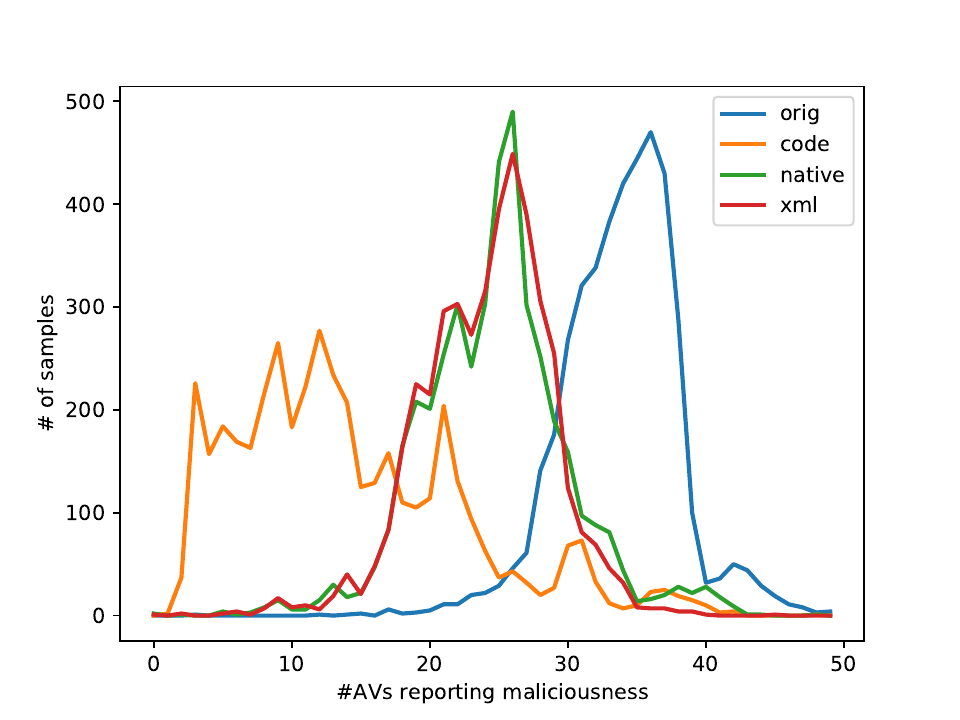}}
		\caption{Dist. of malware vs. \# detected AVs}
		\label{fig:malware-prune}
	\end{subfigure}%
	\caption{Distribution for apps in terms of the number of detected AVs. The curves in blue represent the original apps, orange denotes the apps pruned with code and green curves shows the distribution for the apps pruned by native code, and red for the apps pruned by xml.}\label{fig:prune}
	\vspace{-3mm}
\end{figure*}

\begin{figure*}[t]
	\centering
	\begin{minipage}[t]{0.32\textwidth}
		\centering
		{\includegraphics[width=1.0\textwidth]{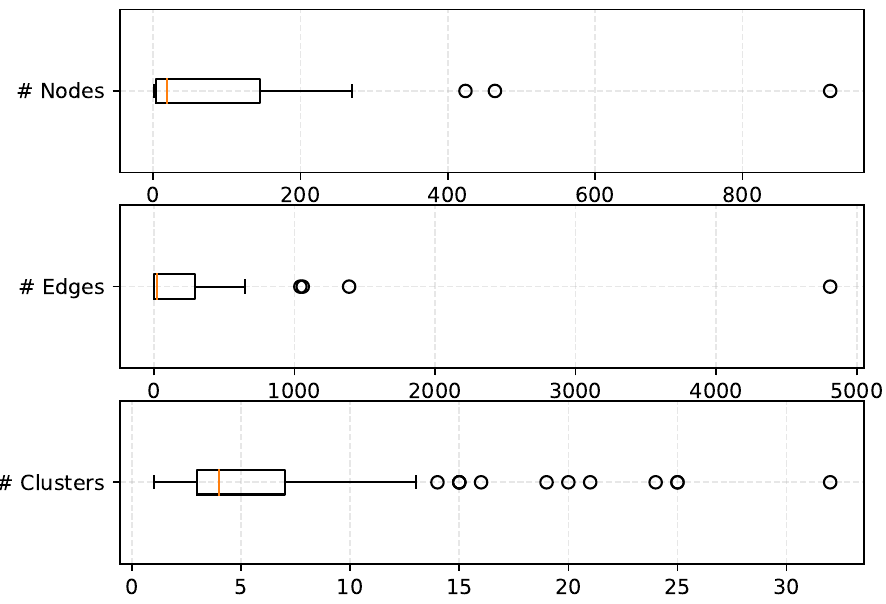}}
		\caption{Statistics of code modularization} 
		\label{fig:modular-cluster}
	\end{minipage}
	\hfill
	\begin{minipage}[t]{0.32\textwidth}
		\centering
		{\includegraphics[width=1.0\textwidth]{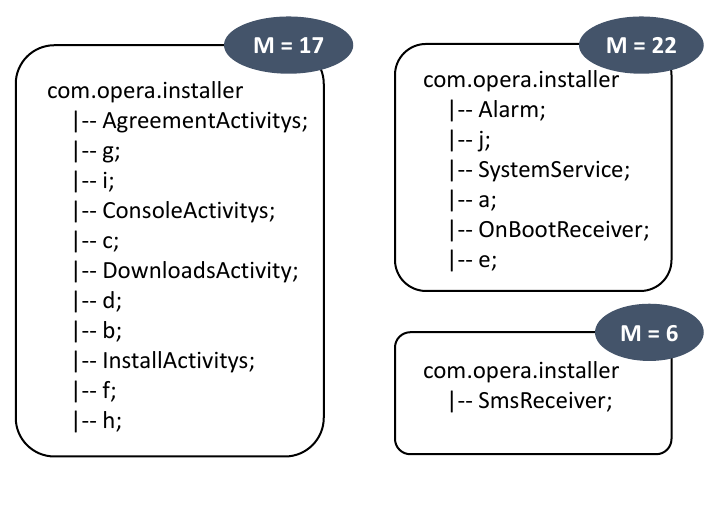}}
		\caption{Modules in the app and their maliciousness with the top-right no.} 
		\label{fig:modular-example}
	\end{minipage}
	\hfill
	\begin{minipage}[t]{0.32\textwidth}
		\centering
		{\includegraphics[width=1.0\textwidth]{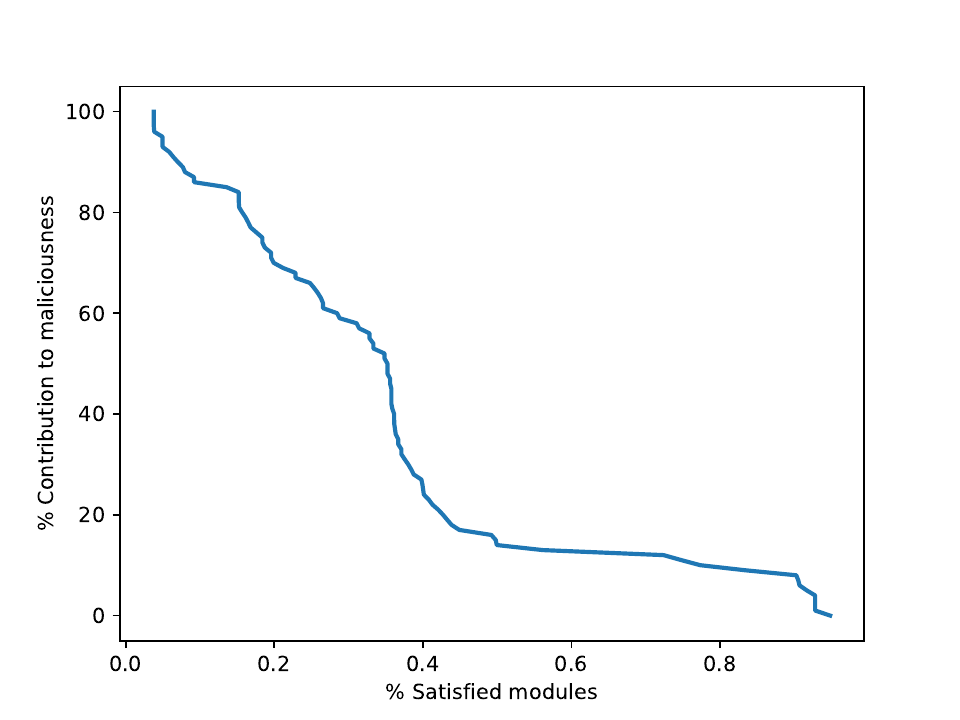}}
		\caption{Contributions of \% of satisfied modules to maliciousness}
		\label{fig:modular-contribution}
	\end{minipage}%
\vspace{-2mm}
\end{figure*}

\subsection{Static Analysis-based Detection (RQ2)}\label{sec:eval:static}

In this study, we propose to employ three transformation techniques--pruning, fusion, and packing, to achieve \emph{function reduction},  \emph{function combination} and \emph{function concealing} of apps, respectively. Together with the changes of detection results, we demystify the usage of static analysis by AVs.

\subsubsection{Function Reduction}\label{sec:eval:static:prune}
Given that APK files typically consist of three semantic modules: Java/Kotlin code, native code~\cite{jnsaf2018ccs}, and XML files, AVs are likely to have extracted features from these modules for detection. To investigate this behavior, we conducted the following experiment.

\vspace{3pt}
\noindent\textbf{Coarse-grained pruning.} We apply \emph{pruning} algorithms to remove these modules from the original apps (Section~\ref{sec:approach:transform}). Using the three strategies, we generate three transformed apps for each original one. In total, we successfully repackage 52,155 apps, each signed with a new certificate. We denote the pruned apps by $P_1$, $P_2$, and $P_3$, corresponding to the removal of Java/Kotlin code, native code, and XML, respectively. Despite potential loss of functionality, observing the detection results shift caused by pruning remains valuable. 

Figure~\ref{fig:prune} shows the distributions of apps as per the number of detected AVs. Besides the distribution of the entire dataset, we also plot the changes of distributions for malware in Figure~\ref{fig:malware-prune}, and grayware in Figure~\ref{fig:grayware-prune}.

\vspace{3pt}
\noindent\textbf{Result Analysis.} Obviously, the maliciousness of $P_1$, $P_2$ and $P_3$ are significantly lower than the original apps. 
The average drops are 53.5\%, 25.2\% and 23.5\%, respectively. 
These drops are more remarkable in malware (\ie, 58.1\%, 28.7\%, 27.0\%) and grayware (\ie, 69.6\%, 28.8\%, 30.7\%). Since the majority of malware has their malicious code in Java/Kotlin, when we remove Java/Kotlin code, malicious behaviors are likely removed accordingly. 
Therefore, the number of detected AVs exhibit a salient decline. 

We examine the effect of pruning on individual AVs, focusing on 32 AVs capable of identifying at least 50\% of malware samples. We compute their maliciousness flips, particularly positive flips, and find that, on average, positive flips occur at rates of 56.7\% for code-pruned apps, 25.2\% for native-pruned apps, and 24.2\% for xml-pruned apps. This is reasonable as Java/Kotlin code is considered the most semantically rich part of an app.

Surprisingly, we identify five AVs with consistently high flip rates across all types of pruned apps: Alibaba (99.98\%, 99.98\%, 99.98\%), MAX (96.0\%, 99.0\%, 95.8\%), McAfee-GW-Edition (93.5\%, 98.3\%, 93.3\%), Jiangmin (89.1\%, 91.6\%, 88.3\%), and Symantec (90.5\%, 76.0\%, 75.8\%). This could be due to these AVs aggregating features from code, native, and XML files into a combined feature for detection. If any part of this feature is removed, detection is significantly affected. This suggests a potential evasion approach by modifying non-critical parts like XML files, which could be effective against these susceptible AVs.

\begin{finding}\label{find:static-prune}
	It is experimentally verified that Java/Kotlin code is the main area to collect features for malware detection. However, we found different use practices such as a combination of features and merely relying on single type of features with regards to a single AV.
\end{finding}

		

\noindent\textbf{Fine-grained pruning.} 
To explore the specific locations of malicious code that attract AVs' attention, we propose fine-grained pruning for Android apps. Typically, AVs combine compromising features from different class files to identify malware, which are semantically related in aspects such as control flow, data flow, and caller-callee relationships. Thus, we aim to separate apps into independent modules and test their maliciousness.
Due to the large number of methods in an app, analyzing at the method granularity is ineffective and inaccurate. Instead, we aim to determine relationships between classes and form clusters accordingly. We develop a script to identify Inter-Component Communications (ICCs) and call graphs in an app using \textsc{IC3-DIALDroid}~\cite{dialdroid2017asiaccs}, ignoring Android APIs, external classes, and resource-related classes. Classes with connections are aggregated into clusters.

Figure~\ref{fig:modular-cluster} shows the percentiles of an app with the number of classes, edges, and formed clusters, while Figure~\ref{fig:modular-example} provides a modularization example for an app. For each group of classes, we generate corresponding pruned apps by eliminating actionable code in other classes. We select the top 150 malware samples with at least 30 reporting AVs for pruning, resulting in 891 fine-grained pruned apps. We stress-test the original 150 apps and 891 fine-grained pruned apps using Monkey, finding that 21 apps execute similarly to their originals, while the rest experience more crashes due to pruning.

\noindent\textbf{Result Analysis.} 
We upload these apps to \virustotal and collect 891 security reports at the first snapshot. We compare the scanning results with that of the original apps (to mitigate repackaging's influence, we repackage the original apps and get their results). For Figure~\ref{fig:modular-example}, we create three pruned apps for the three modules. The original app has $29$ degrees of maliciousness, and the app that eliminate all code can only has $6$ degrees. Maliciousness degrees for these three modules are 17, 21, and 6, respectively. It implies that AVs have obtained more features from the second module. 

In addition, we plot Figure~\ref{fig:modular-contribution} to illustrate the relationship between the contributions to maliciousness with the number of modules. It closes to the Pareto principle that 20\% of modules in an app have contributed around 70\% of maliciousness.
	By grouping code in terms of call relationship, we can identify the most important modules considered by AVs. It can benefit future research on malicious code locating, and interpretation on malware detection results.


\subsubsection{Function Combination}\label{sec:eval:report}
\change{In this section, we propose to employ the \emph{fusion} transformation to merge two samples, combining multiple malicious functions into one app, and further infer how AVs detect the combined malware.}

\vspace{3pt}
\noindent\textbf{Experiment Design}. We cluster four sample sub-sets from our app corpus. In particular, $D_{m}$ contains all malware samples in \genome, \drebin, and \virusshare. $D_{g}$ is the grayware set as Table~\ref{tab:data}. $D_{a}$ is a list of ANVA set, and $D_{b}$ is the sub-set of the wild apps that are not labeled as malware by any AV. The relational behind these combinations is that we intend to investigate how AVs report a combined malware sample by fusing two apps of different maliciousness degrees.  
Additionally, the malicious degress of apps are: $M(malware) > M(grayware) > M(ANVA)$ $\approx$ $M(benign)$ 
As such, we select 3,000 samples in each sub-set, fuse any two sets and obtain 20,305 fused apps.

Given two fused samples $u$ and $v$, $\mathcal{L}(u)$ and $\mathcal{L}(v)$ are the corresponding detection results, respectively. The fusion app is denoted as $w$, and its result $\mathcal{L}(w)$. Inspired by~\cite{avclass2016raid}, we perform suffix removal, tokenization, token filtering, and alias replacement to determine the normalized name for malware samples. 
In this manner, we can determine whether $\mathcal{L}(w)~\approx~\mathcal{L}{(u)}$. 

Upon individual investigation of AVs, AVG, Avast, and Kaspersky demonstrate the greatest likelihood of maintaining original labels when a non-malware sample is merged in. On average, AVG retains the original labels 84.8\% of the time, suggesting its propensity to report known malware features. Avast and Kaspersky exhibit probabilities of 84.6\% and 83.0\%, respectively.
Conversely, some AVs on \virustotal effectively detect malicious features in fused apps. For instance, in the scenario of inserting non-malware code into malware, Alibaba fails to identify 92.1\% of malware samples initially. MAX and Jiangmin exhibit probabilities of 76.2\% and 68.5\%, respectively, suggesting a tendency to label them as non-malware.
In a reanalysis snapshot, Alibaba's percentage drops to 3.4\%, while Jiangmin maintains a high detection rate.

By comparing the results fused by malware with malware, ANVA, and benign apps separately, we can prioritize the types of malware that are more easily detected by AVs. Among the apps labeled differently from one of the original apps, 53.35\% exhibit label changes primarily due to fusion, excluding interference from signatures.
For Avira, 96.6\% of fused apps labeled as ``Malmix'' when one of the original apps is labeled as such, while 77.85\% are labeled as ``BaseBridge'' (including ``BaseBrid'') when one of the original apps is labeled as such. Similar trends are observed for F-Secure (100\% for ``Trojan:Android/IconoSys'', 89.96\% for ``DroidKungFu'', and 83.50\% for ``Trojan:Android/FakeBattScar''), AhnLab-V3 (88.89\% for ``Trojan/Android.Bankun'' and 83.54\% for ``Trojan/Android.BaseBridge''), etc., suggesting that certain features of these malicious types are more easily detected by the corresponding engine or have higher weight in their features.
Conversely, Ikarus exhibits difficulty in detecting apps labeled as ``Trojan-Dropper.AndroidOS.Shedun,'' with 93.03\% changing to another type when fused with other apps, and 6.8\% differing from both original apps. Alibaba also reports the other app's label in 93.41\% of cases when fused with ``TrojanDropper:Android/Shedun.''




\begin{finding}
\xiu{
The majority of malicious features can be well captured, but engines have different detection capabilities for various malware families, especially at the first snapshot. Attackers can exploit this weakened effect to make a simple yet effective evasion to specific AVs.
}
\end{finding}

\begin{table}
	\scriptsize
	\centering
	\caption{AVs with largest positive flip rates and negative flip rates. Here, ``(n=v)'' means the noise of re-signing and the flip rate is $v$ for resigning w/o packing.}\label{tab:packing}
	\begin{tabular}{cc|cc} \toprule
		\textbf{AV} & \textbf{PFR (\%)} & \textbf{AV} & \textbf{NFR (\%)} \\ \midrule
		Alibaba & 99.9 (n=97.7) & K7GW & 100 (n=36.0) \\ 
		Tencent & 99.3 (n=6.6) & ESET-NOD32 & 99.9 (n=0.1)  \\
		AegisLab & 98.0 (n=9.7) & Ikarus & 96.7 (n=0.3) \\
		SymantecMobil. & 97.2 (n=4.6) & Trustlook & \cellcolor{gray} 77.3 (n=86.9) \\
		MAX & 95.6 (n=55.7) & Microsoft & 73.2 (n=11.6)  \\
		Avast-Mobile & 92.7 (n=16.6) & Fortinet & 9.4 (n=6.5)  \\
		Zillya & 91.2 (n=76.2) & TrendMicro-H.C. & \cellcolor{gray} 7.1 (n=8.0)  \\
		Zoner & 91.1 (n=47.0) & Sangfor & \cellcolor{gray} 4.3 (n=8.0) \\
		Qihoo-360 & 90.0 (n=8.5) & F-Secure & \cellcolor{gray} 0.6 (n=1.9) \\
		F-Prot & 89.7 (n=2.1) & Rising & 0.4 (p=0.0) \\	\bottomrule
	\end{tabular}
\vspace{-3mm}
\end{table}
\subsubsection{Function Concealing}\label{sec:eval:static:pack}
We apply packing to conceal functions in apps and assess whether AVs can effectively detect packed apps based on \cite{bangle-code}.
\change{We obtain 15,595 packed apps and their detection results, and present 10 AVs with the largest positive flip rates and 10 AVs with the largest negative flip rates in Table~\ref{tab:packing}. } 
\change{As highlighted in the table, there are four AVs whose negative flip rates are higher via re-signing, i.e., $NFR(Trustlook, \{re-signing\}) > NFR(Trustlook, \{re-signing, packing\})$.}
\change{Since static analysis cannot identify the genuine code from packed apps, many AVs just return labels showing encrypted apps such as ``PUA: Win32/Presenoker'', ``Adware (005487961)'', ``\seqsplit{Android/Packed.Jiagu.E}'', ``\seqsplit{PUA.AndroidOS.Jiagu}''. We found the following exceptions during data analysis.}

\begin{itemize}[leftmargin=*]
	\item \textbf{Incomplete file scanning.} Some AVs, like K7GW, ESET-NOD32 and Microsoft, tend to label malware and benign apps as packed apps. However, Ikarus can correctly recognize all packed samples of the ``BaseBridge'' malware by examining the malicious payload in unencrypted folders like ``assets'' or ``raw.'' It applies to other AVs like ESET-NOD32, SymantecMobileInsight, and Fortinet. However, Ikarus fails to detect payload stored in the ``lib'' folders in the Dowgin malware, indicating incomplete scanning by some AVs.
	\item \textbf{Deficiency in decompressing payload.} Usually, AV engines have an ability of detecting compressed malicious code. However, we found some malicious payloads that are compressed into multiple parts (MD5: \seqsplit{b0d7e14582d58fa6cdacaae65f7b82aa}), which can easily bypass the majority of AVs.
	\item \textbf{Difference in machine learning.} MAX and Trustlook are both machine learning-based antivirus engines, which compute the probabilities of files being malware. However, these two engines react very differently in front of packed apps. MAX has a 95.6\% positive flip rate while Trustlook has a 77.3\% negative flip rate. Because packing mainly conceals the functions in Java/Kotlin code, while reserving the semantics in configure and resource files, the difference indicates that MAX largely relies on harvesting features in code for classification while overlooking the semantics in AndroidManifest.xml or other files.
\end{itemize}

\begin{finding}
Although AVs can raise a warning for packed apps, most of them cannot extract the genuine code in the shell and thereby hardly identify malware or benign apps. Additionally, we identify several weaknesses of specific AVs as described above, for example, incomplete file scanning and deficiency in decompressing payload. 
\end{finding}

 

\subsection{Dynamic Behavioral Analysis (RQ3)}\label{sec:eval:dynamic}
Static analysis-based malware detection effectively identifies large-scale malware but suffers from weaknesses. Malware authors develop evasion techniques, such as obfuscation or concealment, degrading antivirus engine performance. Additionally, malware may not execute as seen, and specific execution paths may never occur during runtime. Dynamic analysis supplements static analysis against malware, with \virustotal equipped with eight sandboxes for security analysis. Sandboxes install and run apps for behavioral analysis, with two (Dr. Web and Tencent) incorporating AVs for real-time malware labeling. We conduct experiments to evaluate contemporary antivirus software's ability.

\subsubsection{Overview of Dynamic Analysis} We randomly select 2000 malware samples as payloads and create a proxy app that downloads them at runtime. If the proxy app is detected as malware, it confirms that antivirus software has dynamically run the host app and checked the maliciousness of the payload. To obtain evidence of dynamic-based detection, we set up a publicly available tracking website to host our selected malware. Upon launching the proxy app and downloading payloads, the tracking website script records connection information, including client user-agent, IP address, visit time, fetched malware samples, etc.

\begin{figure}
	\centering
	\includegraphics[width=0.5\textwidth]{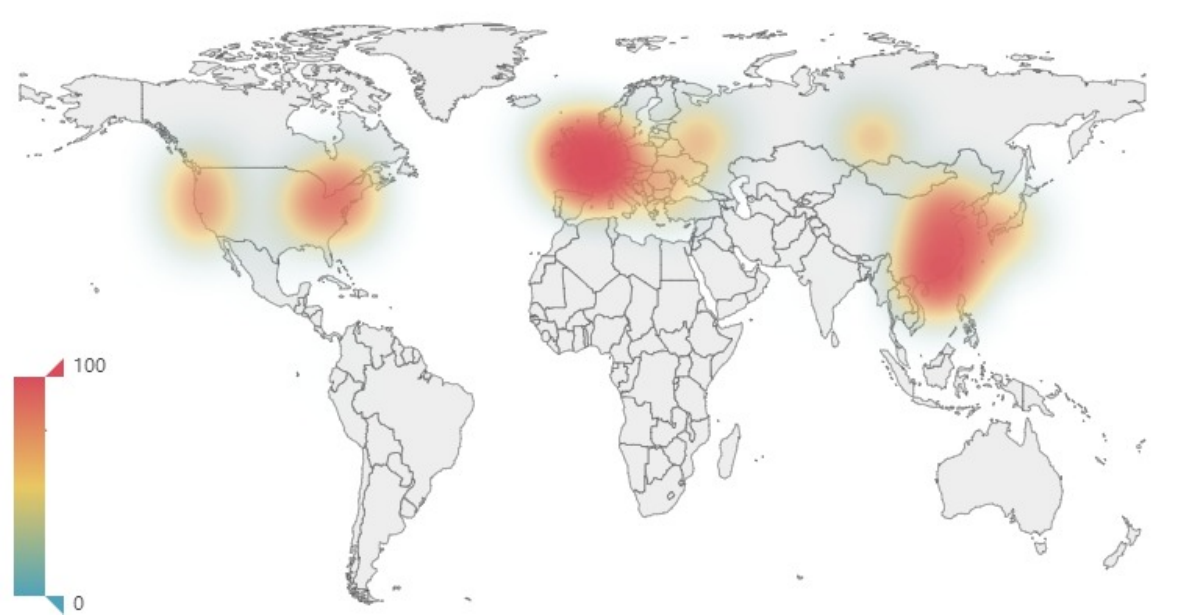}
	\caption{Heatmap of servers that perform dynamic analysis on tested apps}\label{fig:heatmap2}
 \vspace{-3mm}
\end{figure}

We collected \emph{7}-day reports for the apps, totaling 50,543 requests from 251 distinct IP addresses. To identify real scanning entities among these IP addresses, we crawled configuration information for each IP address using IP-API\footnote{https://ip-api.com/}, including continent, country, city, exact location, ISP, etc. We grouped IP addresses of one scanning entity based on the assumption that two IP addresses belong to the same network with the subnet mask 255.255.255.0, and their configuration information is identical except for the IP address. This yielded 49 individual scanning entities. A heatmap in Figure~\ref{fig:heatmap2} shows network flow, with Roubaix (19.5\%), Strasbourg (18.7\%), Guangzhou (13.3\%), Santa Clara (11.6\%), and Ashburn (5.5\%) having the largest flow. On average, each entity dynamically ran 259 samples (25.9\%) over the 7-day period. Most dynamic scans (90.2\%) were completed within 24 hours. Besides, we detected 92 requests that were not made by our proxy apps, suggesting that AVs also attempt to traverse all possibly related URLs.

\begin{finding}\label{find:url}
	We identified 49 dynamic analysis entities within the 7-day network flow tracking, and some of them also perform a fuzzy search by mutating URLs to dig out more connections. 
\end{finding}



\noindent\textbf{Label dynamics}. We track the detection results of these disguised apps by \virustotal for 7 days, and also get the reanalysis snapshot for these apps.
Surprisingly, in the 8,000 security reports, only \emph{Ikarus} is able to detect the proxy app from the first day (We also confirm that the URL has not been marked as malicious).
Although two engines from Dr. Web and Tencent have carried on their dynamic analysis within sandboxes, the corresponding engines hosted in \virustotal fail to label these malicious apps. 
Take as an example the proxy app that loads malware (MD5:\seqsplit{9d795006c4fff9c61d460d7e30de364d}), the label returned by Ikarus  ``\texttt{\seqsplit{Trojan.AndroidOS.Agent}}''. Compared with detection results at the same day, the label given by Ikarus is ``\texttt{\seqsplit{AdWare.AndroidOS.DroidRooter}}'', which is totally different. 
Since our proxy apps have never been seen by these antivirus engines, it is thus concluded that Ikarus has performed a dynamic analysis to the new sample and label the newly detected samples with consideration of similar malware in the database.

\begin{finding}\label{find:dynamic-result}
Through the 7-day tracking on detection results, only one AV can successfully detect the maliciousness of our proxy apps. Even the AVs whose sandboxes are integrated by \virustotal, fail to detect or synchronize with a correct detection result.
\end{finding}



\subsubsection{Measurement of Analysis Capability}
It is learned from the last experiment that some AVs and sandboxes in \virustotal are performing dynamic analysis of uploaded apps. To further measure their capabilities, we craft a number of apps in this section. 
Specifically, we inject \emph{logic bombs}~\cite{triggerscope2016sp} into malware samples to hinder dynamic-based analysis. 

For simplicity, we do not take into account infinite code constraints like \texttt{if-else} statements. 
Instead, we employ SMS, location and broadcast events as a trigger in ~\cite{triggerscope2016sp} for malicious payload downloading. First, we rank the occurrences of broadcast events from 11,566 malware samples and determine the top 10 in Table~\ref{tab:actions} with their occurrence rates amongst malware. 
Additionally, we list the location event separately at the bottom since this event is not captured by a \texttt{BroadcastReceiver}.
In this way, we create 10 proxy apps from H1 to H10 with a trigger. We implement the events ``\texttt{\seqsplit{\small android.provider.Telephony.SMS\_RECEIVED}}'' and ``\texttt{\seqsplit{\small android.provider.Telephony.SMS\_DELIVER}}'' into one app since both of them are for an incoming SMS message. 

To conduct our measurement, we design the proxy app to notify our tracking server upon startup. Upon receiving an acceptable event, the app initiates the download of malicious payloads, leaving a trace on our server. To mitigate interference, we obfuscate our server addresses with meaningless strings, as suggested by Finding \ref{find:url}. 
We create 100 samples for each type of proxy app, and each of them will download a different malware samples. Totally, we get 1000 proxy apps and then upload them into \virustotal. With a 7-day observation, we receive 1212 requests from 2 IP addresses. In addition, we collect in total 2000 malware detection reports at the first and reanalysis snapshots.

We observe one server from Bitdefender and another in Dublin. The requests originate from four proxy apps: H2, H3, H4, and H5. On average, H4 triggers successfully in 0.83s with a 100\% success rate, H3 in 0.67s with a 95.9\% success rate, H2 in 14.1s with a 95.8\% success rate, and H5 in 0.56s with a 78.5\% success rate. 
Surprisingly, we identify an exception, \ie, H1, whose download requests come first but the initial requests are never captured. 
This can be largely attributed to a testing sequence happening in a sandbox: install the proxy app, send a broadcast event of BOOT\_COMPLETED. However, the proxy app is never actually started. 
The remaining five types show no evidence of dynamic analysis. Examination of detection reports from two snapshots reveals no negative results for proxy apps. While numerous requests were captured, most are likely due to static analysis, as URLs are extracted statically, and AVs or other tools generate visit requests. 

\begin{finding}\label{find:dynamic-cap}
	Sandboxes that dynamically analyze Android apps have integrated a considerable number of triggering techniques such as broadcast events, SMS service, etc. The triggering success rate is up to 93.2\% on average from the captured evidence. In addition, urls contained in an app may also be examined without executing apps.
\end{finding}
\begin{table}
	\centering
	\scriptsize
	\caption{Top 10 broadcast events that are frequently received by malware, and the additional event-location. ``Perm'' indicates whether a permission is required.}\label{tab:actions}
	\resizebox{1\linewidth}{!}{
		\begin{tabular}{clcl} \toprule
			\textbf{Host} & \textbf{Action} & \textbf{Perm?}&\textbf{Per. (\%)} \\ \midrule
		H1 & 	android.intent.action.BOOT\_COMPLETED & \ding{55} & 58.2  \\
		H2 & 	android.net.conn.CONNECTIVITY\_CHANGE & \ding{51}  &39.8  \\
		H3 & 	android.intent.action.USER\_PRESENT  & \ding{55} &  34.4 \\
		H4 & 	android.provider.Telephony.SMS\_RECEIVED & \ding{51} &  28.3 \\
		H5 & 	android.intent.action.PACKAGE\_ADDED &  \ding{55} & 19.7 \\
		H6 & 	android.intent.action.PACKAGE\_REMOVED & \ding{51}  &  18.2 \\
		H7 & 	android.provider.Telephony.SMS\_DELIVER &\ding{51}  & 14.1 \\
		H8 & 	android.intent.action.ACTION\_POWER\_CONNECTED & \ding{55} & 11.9 \\
		H9 & 	android.intent.action.ACTION\_POWER\_DISCONNECTED & \ding{55} & 8.7 \\
		H10 & 	android.intent.action.SCREEN\_ON  & \ding{55}  & 8.5 \\ \midrule \midrule
		H11 &  [LOCATION CHANGE EVENT] & \ding{51} & - \\ \bottomrule
			
		\end{tabular}
	}
\end{table}

\section{Threats to Validity}

The threats to the experiments and results primarily arise from three aspects. First, the determinacy of detection results is uncertain. Since AVs operate as fuzzy systems and may return unexpected results, as noted in previous research, this inaccuracy can negatively impact the analysis results and conclusions drawn from them. Second, approximately 3.6\% of apps cannot be successfully unsigned and re-signed, rendering them unusable for other transformations. However, this accounts for a relatively small portion, allowing our experimental apps to represent the majority. In dynamic analysis, although we set up a tracking server to monitor app execution, we are unable to capture requests from apps in an isolated sandbox, limiting our ability to confirm the existence of dynamic analysis rather than its absence. Lastly, during the diff analysis, we consistently select results from an AV with identical versions and update dates. However, this may not always be feasible, leading us to sometimes use results produced on adjacent or relatively close dates.

\section{Related Work}\label{sec:related}


\vspace{3pt}
\noindent\textbf{Evaluation of antivirus engines}.
AVs are evaluated with transformed apps in prior studies. For example,
\textsc{DroidChameleon}~\cite{droidchameleon2013asiaccs} is an approach to obfuscate app code and verify whether AVs are still capable of recognizing malware. 
\textsc{Mystique}~\cite{asiaccs2016mystique,mystiques2017} have employed a generic algorithm to evolve malware by adding more features and evaluate anti-malware tools. These transformations are all compound and coarse-grained, where AVs' failures may not be only attributed to code obfuscation but also the broken of app integrity.
Cai and Yap \cite{cai2016codaspy} compare the detection results between malware and obfuscated malware, and unveil AVs' detection logic. 
Our study has enriched transformation strategies and achieved a similar but more accurate goal with a much larger dataset.
Studies \cite{preda2017obfuscation,empirical2018icse} are dedicated to a large-scale empirical study on the effects of code obfuscation against antivirus software. They compare the detection results by 61 antivirus engines and measure the performance of obfuscation strategies and tools. 
Huang~\etal~\cite{dilemmas2014secure} conduct a study to explore the issues of antivirus engines in modern mobile platforms. 
Murali \etal~\cite{MURALI2023120092} propose an evolutionary algorithm to generate malware antigen, which can be used to construct variants for a given source malware.
As for dynamic analysis-based antivirus engines, \textsc{AVLeak}~\cite{avleak2016woot} can fingerprint an emulator via black box testing for privacy stealing. 
Quarta \etal~\cite{av2018dimva} propose a blackbox method to explore if an AV implements emulation, static unpacking and heuristics matching.   
Additionally, some studies propose sandbox evasion techniques to test antivirus engines. Yokoyama~\etal\cite{yokoyama2016raid} fingerprint the unique features of sandboxes and train a classifier to tell apart a sandbox and a real system. However, the features of a sandbox can be replaced with more realistic values, in case of malware evasion. Miramirkhani~\cite{miramirkhani2017sp} propose Wear-and-Tear artifacts occurring in a sandbox. Consequently, malware can effectively determine their running environment and evade dynamic analysis.
\emph{Apart from prior research, our study leverages the dynamics of malware labels to determine whether AVs have anticipated detection behaviors. By transforming Android malware and benign apps, we can learn how AVs react to these tracable changes and measure their performance. }

\noindent\textbf{Malware Labelling}. Reports returned by online scanning services may be inconsistent and confusing. Therefore, the works~\cite{virustotal2019imc,virustotal2020usenix} take long-time snapshots of \virustotal reports of malware samples, and unveil the dynamics of malware labels. 
Malware types are varying across different antivirus engines. To clarify this confusion, the works~\cite{avmeter2014dimva,avclass2016raid,euphony17msr} learn the contextual information contained in the reported malware names, and reason out the most convincing type for malware. 
Wei \etal employ a lightweight static analysis to measure the similarity with well-labeled malware and create the AMD dataset~\cite{amd2017dimva}. 
There is a line of research on identifying noisy training data in machine learning-based malware detection. Xu \etal\cite{ndss2021label} propose a differential training to obtain intermediate states of two identical classification models as noise-detection features. Then a set of outlier detection algorithms are applied to identify noisy data. 
Wang \etal\cite{malwhiteout,wang24ase} propose MalWhiteout, a confidence learning approach to identify noises in malware labels, and further improve machine learning-based malware detection. It incorporates ensemble learning and inter-app relation to reduce false positives of label noises.
\emph{Compared to prior work, our approach takes into account the dynamics of malware labels via massive app transformations, and further measure the robustness of antivirus engines. The results can help to understand the working mechanism of these engines and improve their abilities in malware detection.}
\section{Conclusion}\label{sec:concl}
We propose a data-driven approach to explore blackbox AVs and infer the detection strategies in AVs by transforming Android apps. We first put forward an interaction model to simulate the behaviors of blackbox AVs. 
Guided by the model, we create a large corpus of apps consisting of crawled apps from multiple sources, and transformed apps by six transformations. 
971K security reports are collected from \virustotal. Then we conduct a comprehensive measurement from three aspects: signature-based, static analysis-based detection, and dynamic analysis. Last, we draw 6 findings that are useful in AV-involved research and practices. 

\section*{Acknowledgment}
Thanks for all the anonymous reviewers and their constructive comments. The IIE authors are partially supported by the Strategic Priority Research Program of Chinese Academy of Sciences under Grant No. XDB0690100.

\pagebreak
\balance
\bibliographystyle{ACM-Reference-Format}
\bibliography{main}

\end{document}